\documentclass[twocolumn,pra,amsmath,amssymb]{revtex4}

\usepackage{graphicx}
\usepackage{enumerate}

\newtheorem{thm}{Theorem}
\newtheorem{lem}[thm]{Lemma}

\begin{document}
\title{The structure of Bell inequalities}
\author{G\"{u}nter Schachner}
 \email{gus@q-te.com}
 \noaffiliation
\date{\today}

\begin{abstract}
In this paper we present an analysis of the structure of Bell inequalities, mainly for the case of 
$N$ qubits with two observables each. We show that these inequalities are related to 
\emph{Hadamard matrices} and define \emph{Bell polynomials} (in one variable)
as an additional tool.
With these aids we raise several conditions the coefficients of Bell inequalities
must satisfy, and recursively generate the whole set of inequalities starting from $N=1$.
Moreover, we prove some characteristic features of this set, such as 
that most of the inequalities contain all expectation values under consideration. 
Finally, we show how the presented results can be used to construct Bell 
inequalities with certain properties. An outlook on further research topics concludes the paper.
\end{abstract}

\maketitle

\section{Introduction} \label{S:intro}
In 1964, J.\,S.~Bell published his now-famous paper \cite{bell:64}, which 
revolutionized the foundations of quantum mechanics and our view of nature.
After decades of discussion Bell demonstrated that it was 
possible to decide \emph{experimentally} whether the EPR argument \cite{epr} was correct.
The central part of this paper was an inequality assigned to a specific experimental 
setup. The expectation values of measurements had to satisfy this inequality so 
that a local realistic model could be applied. Even for more complex quantum systems 
there always exists a set of linear constraints that serve this purpose. Such constraints 
are today called \emph{Bell inequalities}.

\begin{figure*}%
\includegraphics{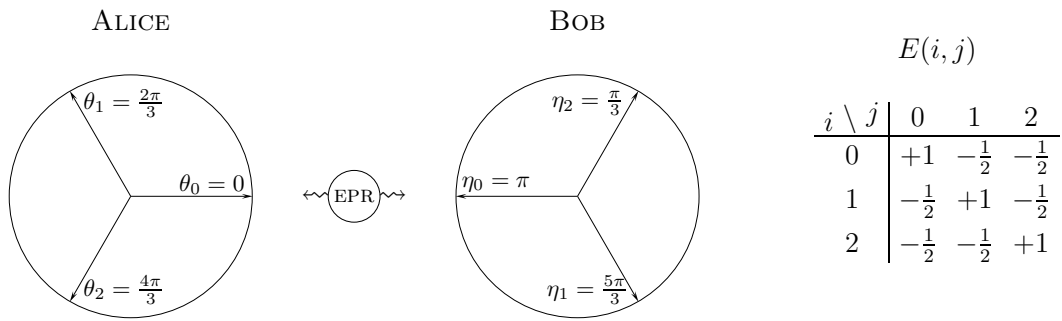}
\caption{The experimental setup described in the text and the corresponding expectation values $E(i,j)$.}
\label{F:setup}
\end{figure*}

Let us consider two observers, Alice and Bob, who perform measurements 
on \emph{qubits} (the simplest quantum systems, having two possible outcomes). 
We assume that each of them has the choice of measuring
one out of three observables. According to D.~Bohm's variant \cite{bohm:51} of the EPR 
experiment \cite{epr}, these measurements can e.g.\ be spin measurements on particles of 
spin $\frac{1}{2}$, where Alice and Bob have the option of measuring at angles
\begin{equation} \label{E:theta}
\theta_0 = 0,\ \theta_1 = \frac{2\pi}{3},\ \theta_2 = \frac{4\pi}{3}
\end{equation}
and
\begin{equation} \label{E:eta}
\eta_0 = \pi,\ \eta_1 = \frac{5\pi}{3},\ \eta_2 = \frac{\pi}{3},
\end{equation}
respectively (see Fig.~\ref{F:setup}). If we assign values to the possible outcomes, 
say $+1$ to ``spin up'' and $-1$ to ``spin down,'' we are able to examine 
\emph{expectation values} over many repetitions of the experiment. 
Let $E_1(i)$ and $E_2(j)$ denote the expectation values
of measurements at angles $\theta_i$ and $\eta_j$, respectively, and let
$E(i,j)$ denote the expectation value of the product of the assigned values.
If Alice and Bob use an entangled pair of particles from a source in 
singlet state that shows no preference for any specific direction, we get
\begin{equation*}
E_1(i) = E_2(j) = 0 \quad \text{for $0 \leq i,j \leq 2$.}
\end{equation*}
Moreover, quantum mechanical calculation \cite{peres:95} yields
\begin{equation*}E(i,j) = -\cos \bigl(\theta_i - \eta_j\bigr) = 
\begin{cases}
+1          & \text{\ if $i=j$}\\
-\frac{1}{2} & \text{\ otherwise},
\end{cases}
\end{equation*}
in perfect accordance with empirical data. Thus, if the choices
made by Alice and Bob are independent of one another and each angle is chosen with
equal frequency, we get the following properties:
\begin{enumerate}
\item If $i=j$, then the same values are measured.
\item The overall expectation value of products is zero. \label{I:exp_zero}
\end{enumerate}
There is no way of explaining this behavior in terms of a local realistic model, 
where the outcomes of Alice's measurements are \emph{independent} of Bob's measurements 
and vice versa. 

\medskip\noindent\textit{Remark.}
Property \ref{I:exp_zero} is still satisfied if one or both of the apparatures are tilted 
by some arbitrary degree, because for any $\phi$ there is
\begin{equation*}
\cos\Bigl(\frac{\pi}{3}+\phi\Bigr) + \cos(\pi+\phi) + \cos\Bigl(\frac{5\pi}{3}+\phi\Bigr) = 0.
\end{equation*}
Thus, essentially the same argumentation holds if
\begin{equation*}
\eta_i = \theta_i \quad \text{for $0 \leq i \leq 2$}
\end{equation*}
is used instead of \eqref{E:eta}, which is often the case in the literature 
(e.g.\ in \cite{mermin:85}).

In a general setup there are $N$ observers having the choice of measuring one out of 
$g$ observables, where the outcome of each measurement is $q$-valued.
If we enumerate the observables by $0, 1, \ldots, g-1$, their respective 
choices can be described by a vector
\begin{equation*}
(k_1, k_2, \ldots, k_N)
\end{equation*}
with $0 \leq k_i \leq g-1$ for $1 \leq i \leq N$.
We can interprete this vector as the \emph{$g$-adic expansion} of an integer
\begin{equation} \label{E:g-adic}
k = (k_1, k_2, \ldots, k_N)_g = \sum_{i=1}^N k_i g ^{N-i}
\end{equation}
with $0 \leq k \leq g^N-1$. (If there are less than $g$ observables at some sites,
$k$ simply does not take all possible values.)

In a local realistic model each observable is a random variable $A_i(k_i)$ 
in its own right and is \emph{independent} of the choices 
$k_j$ for $j \neq i$ at other sites. We will denote the corresponding expectation values 
of the product of these variables by
\begin{equation} \label{E:expect}
E(k) := E(k_1,\ldots,k_N) = \Bigl\langle A_1(k_1) \cdots A_N(k_N) \Bigr\rangle,
\end{equation}
which must in any local realistic model satisfy a set of Bell inequalities.

As the systems grow, the number and complexity of these inequalities increase dramatically.
In fact, in \cite{pitowsky:89} it is shown (in terms of joint probabilities, instead of expectation values) 
that the question of whether a local realistic model can be applied or not
is related to a convex \emph{correlation polytope}.
The experimental results can be explained by a classical probability distribution 
exactly if the corresponding vector of probabilities and joint probabilities lies
inside that polytope. This problem is NP-hard. (For the definition and a survey of NP-hard problems 
see \cite{garey-johnson:79}.) Historically, related problems were already investigated by 
G.~Boole \cite{boole} in the $19^\text{th}$ century. Independently, these problems are of relevance 
in probability theory and related research is going on to this day 
(see also \cite{pitowsky:89a} for a discussion).

\medskip\noindent\textit{Remark.}
In the literature the enumeration of observables usually starts with
$1$ instead of $0$. Therefore, in the literature the expectation value 
\eqref{E:expect} is written as 
\begin{equation*}
E(k_1+1,\ldots,k_N+1)
\end{equation*} 
with $0 \leq k_i \leq g-1$ for $1 \leq i \leq N$. 
We will also pay attention to that convention in this paper, 
referring to it as \emph{traditional notation}.

\section{Bell inequalities} \label{S:bell_equ}
We will now and for the rest of this paper study the case of $N$ qubits with
two observables each, i.e.\ $q=2$ and $g=2$. For that purpose we consider (see also \cite{zukowski-brukner:02})
the product
\begin{equation} \label{E:prod}
P(h_1,\ldots,h_N) := \prod_{i=1}^N \bigl(A_i(0) + h_i A_i(1)\bigr)
\end{equation}
for arbitrary $h_i \in \{-1,1\}$, which can be expanded to
\begin{equation} \label{E:S_h}
\hspace*{-.1cm}P(h_1,\ldots,h_N) = \hspace*{-.85cm} \sum_{(k_1,\ldots,k_N) \in \{0,1\}^N} \hspace*{-.85cm} h_1^{k_1} \cdots h_N^{k_N} A_1(k_1) \cdots A_N(k_N).
\end{equation}
That again defines a random variable, which now also depends on the
nonrandom variables $h_1,\ldots,h_N$. 
For a concrete realization of variables $A_i(k_i)$, there is only one 
choice for the $h_i$'s so that the product \eqref{E:prod} does not vanish, 
in which case each factor is $\pm 2$. Thus, we have
\begin{equation*}
\sum_{(h_1,\ldots,h_N)\in\{-1,1\}^N} \hspace*{-.85cm} P(h_1,\ldots,h_N) = \pm 2^N.
\end{equation*}
Since this sum contains only one nonvanishing term, we also get
\begin{equation} \label{E:S_sum}
\sum_{(h_1,\ldots,h_N)\in\{-1,1\}^N} \hspace*{-.85cm} c(h_1,\ldots,h_N) P(h_1,\ldots,h_N) = \pm 2^N
\end{equation}
for an arbitrary $\pm 1$-valued function $c(h_1,\ldots,h_N)$. The expectation value of that sum must 
therefore lie between $-2^N$ and $2^N$. In order to derive constraints for \eqref{E:expect}, 
we substitute \eqref{E:S_h} in that expression and use linearity of expectation.
With respect to \eqref{E:g-adic} we set
\begin{equation} \label{E:coeff}
a_k := \hspace*{-.5cm}\sum_{(h_1,\ldots, h_N) \in \{-1,1\}^N} \hspace*{-.85cm} h_1^{k_1}\cdots h_N^{k_N} \, c(h_1,\ldots,h_N)
\end{equation}
for $0 \leq k \leq 2^N-1$, and finally get
\begin{equation} \label{E:bell}
\biggl\lvert\sum_{k=0}^{2^N-1} a_k E(k) \biggr\rvert \leq 2^N.
\end{equation}
By choosing all admissible functions $c(h_1,\ldots,h_N)$ in \eqref{E:coeff}, the corresponding
inequalities \eqref{E:bell} represent a complete set of Bell inequalities for the 
experimental setup under consideration \cite{werner-wolf:01,zukowski-brukner:02}. That means
that these inequalities are satisfied exactly if a local realistic model can be applied.
 
\medskip\noindent\textbf{Example}
Let us consider the case $N=2$. By \eqref{E:coeff} we have
\begin{align*}
a_0 &= c(1,1) + c(1,-1) + c(-1,1) + c(-1,-1) \\
a_1 &= c(1,1) - c(1,-1) + c(-1,1) - c(-1,-1) \\
a_2 &= c(1,1) + c(1,-1) - c(-1,1) - c(-1,-1) \\
a_3 &= c(1,1) - c(1,-1) - c(-1,1) + c(-1,-1)
\end{align*}
with an arbitrary $\pm 1$-valued function $c(h_1,h_2)$.
For example, if we choose
\begin{equation*}
c(h_1,h_2) := 1-\frac{(h_1+1)(h_2+1)}{2}
= \begin{cases}
  -1 & \text{if $h_1=h_2=1$}\\
  +1 & \text{otherwise},
  \end{cases}
\end{equation*}
we get
\begin{equation*}
a_0 = 2 \quad \text{and} \quad a_1 = a_2 = a_3 = -2.
\end{equation*}
By \eqref{E:bell} this leads, after division by $2$, to
\begin{equation} \label{E:chsh1}
\lvert E(0) - E(1) - E(2) - E(3) \rvert \leq 2,
\end{equation}
which in traditional notation reads as
\begin{equation} \label{E:chsh2}
\lvert E(1,1) - E(1,2) - E(2,1) - E(2,2) \rvert \leq 2.
\end{equation}
(The transcription from \eqref{E:chsh1} to \eqref{E:chsh2} happens by writing each 
argument in its binary expansion, using exactly $N$ digits, 
and then incrementing each digit by $1$.) By using all $2^4=16$ admissible functions 
$c(h_1,h_2)$, we can easily verify that this inequality is, up to symmetry, the only 
nontrivial case for $N=2$.\hfill\ensuremath{\Diamond}\medskip

Inequality \eqref{E:chsh2} was first derived by 
J.\,F.~Clauser, M.\,A.~Horne, A.~Shimony \& R.\,A.~Holt \cite{chsh},
and we will therefore subsequently refer to it and its symmetric variants 
as \emph{CHSH inequalities}. From now on we will also use the shorthand notation
\begin{equation} \label{E:shorthand}
(a_0, a_1, \ldots, a_{2^N-1})
\end{equation}
for \eqref{E:bell}. This is convenient since we will see that even if \eqref{E:bell} is 
multiplied by an arbitrary nonzero constant we can still calculate its upper bound by plain use of
\eqref{E:shorthand}. (Without such multiplication this is trivial, since in that case 
we only need to take the length of this vector to achieve this bound.)

\section{Hadamard matrices}
The vector $(h_1,\ldots,h_N) \in \{-1,1\}^N$ can also be interpreted as a binary 
expansion with ``digits'' $\pm 1$. Thus, if we use
the substitutions $1 \mapsto 0$ and $-1 \mapsto 1$ in this expansion, we get 
a nonnegative integer
\begin{equation*}
j = (j_1,\ldots,j_N)_2,
\end{equation*}
where $h_i = 1-2j_i$ for $1 \leq i \leq N$. This leads us to write $c_j$ instead of 
$c(h_1,\ldots,h_N)$, which in the previous example means
\begin{equation*}
c_0=c(1,1),\, c_1=c(1,-1),\, c_2=c(-1,1),\, c_3=c(-1,-1)
\end{equation*}
and further
\begin{equation*}
\begin{pmatrix}
a_0\\ a_1\\ a_2\\ a_3
\end{pmatrix}
=
\begin{pmatrix}
1 & \phantom{-}1 & \phantom{-}1 & \phantom{-}1\\ 
1 & -1           & \phantom{-}1 & -1\\
1 & \phantom{-}1 & -1 & -1\\
1 &           -1 & -1 & \phantom{-}1\\
\end{pmatrix}
\begin{pmatrix} 
c_0 \\ c_1 \\ c_2 \\ c_3 
\end{pmatrix}.
\end{equation*}

We will now study the matrices involved in this operation
in general. For that purpose we consider the binary expansions
$j = (j_1,\ldots,j_N)_2$ and $k = (k_1,\ldots,k_N)_2$
and their scalar product over $\text{GF}(2)$, defined as
\begin{equation} \label{E:scalar}
\langle j,k \rangle_2 := \sum_{i=1}^N j_i k_i \pmod{2}.
\end{equation}
(In other words, $\langle j,k \rangle_2 = 1$ if the number of $1$'s in which the 
binary expansions of $j$ and $k$ coincide is odd, and $0$ otherwise.) We observe in
\eqref{E:coeff} that $h_i^{k_i}=-1$ if and only if $h_i = -1$ and $k_i=1$, which means
that the corresponding bits in $j$ and $k$ are both $1$. Therefore, we can
also write \eqref{E:coeff} as
\begin{equation}
a_k = \sum_{j=0}^{2^N-1} (-1)^{\langle j,k \rangle_2} c_j
\end{equation}
with an arbitrary $\pm 1$-valued vector $(c_0, c_1, \ldots, c_{2^N-1})$. 
Thus, if we define the ${2^N \times 2^N}$ matrix $H_{2^N} = (h_{jk})$ by
\begin{equation} \label{E:h_jk}
h_{jk} := (-1)^{\langle j,k \rangle_2}
\end{equation} 
for $0 \leq j,k \leq 2^N-1$, we finally get
\begin{equation} \label{E:matrix}
\begin{pmatrix}
a_0\\ a_1\\ \vdots\\ a_{2^N-1}
\end{pmatrix}
=\ H_{2^N} \begin{pmatrix} c_0\\ c_1\\ \vdots \\ c_{2^N-1} \end{pmatrix}.\\
\end{equation}

\begin{figure*}%
\includegraphics{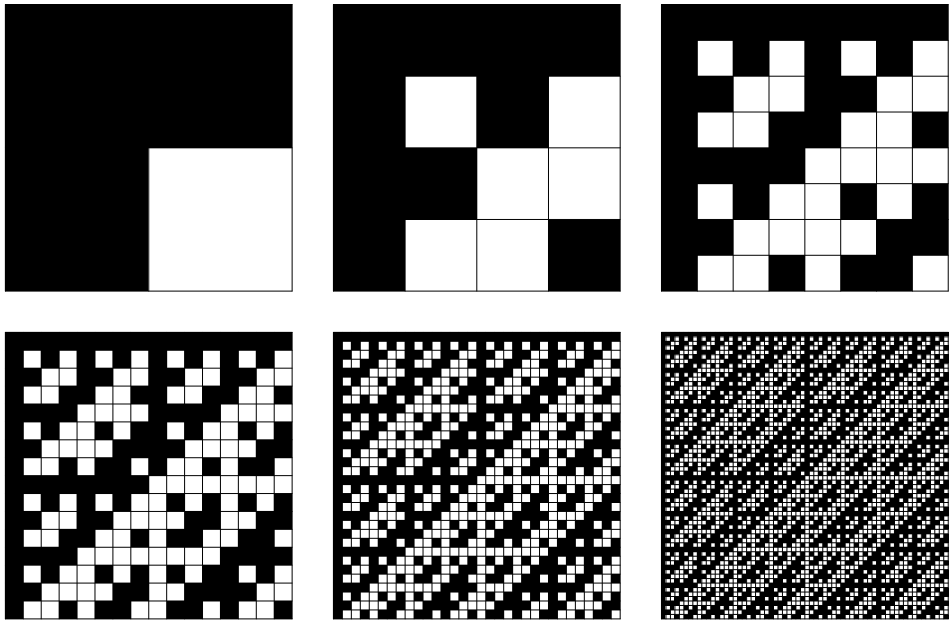}
\caption{The Hadamard matrices $H_{2^N}$ for $1 \leq N \leq 6$. A black square symbolizes $+1$, a 
         white square $-1$.}
\label{F:hadamard}
\end{figure*}

The matrices $H_{2^N}$ constitute a special type of \emph{Hadamard matrices}. 
These are matrices with elements $\pm 1$, where two rows (and columns) differ in 
exactly half of their elements. (This is another way to say that 
Hadamard matrices are orthogonal $\pm 1$-valued matrices.) They have the additional 
property of having \emph{maximal determinants} among all -- even complex -- matrices with 
elements bound by $1$ (in absolute values); namely, if $H_n$ denotes an arbitrary Hadamard 
matrix of dimension $n \times n$, then
\begin{equation*}
\lvert \det H_n \rvert = n^{\frac{n}{2}}.
\end{equation*}
(This property comes as no surprise, since this determinant can be interpreted 
geometrically as the volume of a parallelepiped spanned by the row vectors of 
$H_n$ -- each of length $\sqrt{n}$ --, which is maximal if these vectors are orthogonal.)
Historically, this was also the reason why such matrices were originally studied 
by J.\ Hadamard \cite{hadamard:93}. It is not difficult to prove that for any Hadamard matrix~$H_n$ 
there is $n=1$, $n=2$ or $n=4k$ for some $k \geq 1$ (e.g.\ see \cite{wallis:72}); 
but it is open whether a Hadamard matrix really exists for each such $n$. 
(The first unknown case up to this day is $n = 428$.) For further information on 
Hadamard matrices see \cite{wallis:72, sloane}.

The matrices $H_{2^N}$ can also be derived \cite{wallis:72} via the recursive definition
\begin{align} \label{E:hadamard}
H_1    &= \bigl(1\bigr) \notag \\
H_{2n} &= \begin{pmatrix} H_n & \phantom{-}H_n \\ H_n & -H_n \end{pmatrix} \quad \text{for $n \geq 1$}.
\end{align}
This construction was first given by J.\,J.~Sylvester in the context of tilings \cite{sylvester:67},
26 years before Hadamard studied matrices of that kind. Therefore, these matrices
are called \emph{Sylvester-type} Hadamard matrices. In particular,
they are normalized, i.e.\ the elements of the first row and column are all $1$, 
and symmetric. Using the Kronecker (or tensor) product $\otimes$ for matrices 
$A=(a_{ij})$ and $B$, defined as
\begin{equation*}
A \otimes B = \begin{pmatrix} a_{11}B & a_{12}B & \cdots \\ a_{21}B & a_{22}B & \cdots \\ \vdots & \vdots & \ddots \end{pmatrix},
\end{equation*}
we can write \eqref{E:hadamard} for $n \geq 2$ also as $H_{2n} = H_2 \otimes H_n$.
Therefore we get 
\begin{equation*}
H_{2^N} = \underbrace{H_2 \otimes \cdots \otimes H_2}_N.
\end{equation*}
(The operator $\otimes$ is associative, thus it makes no difference whether
this expression is evaluated from the left or from the right.)
Explicitly written, that means
\begin{gather*}
\begin{aligned}
H_1 &= \bigl(1\bigr) \\
H_2 &= \begin{pmatrix} 1 & \phantom{-}1 \\ 1 & -1 \end{pmatrix} \\
H_4 &= \begin{pmatrix} 1 & \phantom{-}1 & \phantom{-}1 & \phantom{-}1\\ 
                       1 & -1           & \phantom{-}1 & -1\\
                       1 & \phantom{-}1 & -1 & -1\\
                       1 &           -1 & -1 & \phantom{-}1\\
        \end{pmatrix}\\
\end{aligned}\\
\cdots\cdots\cdots\cdots\cdots\cdots\cdots\cdots
\end{gather*}
For a visualization of the matrices $H_{2^N}$ for $1 \leq N \leq 6$ see Fig.~\ref{F:hadamard}.

\medskip\noindent\textit{Remark.}
For matrices of type \eqref{E:hadamard}, there also exist numerous other descriptions;
e.g.\ they can as well be defined as the table of characters for elementary Abelian groups of 
order $2^N$. Moreover, Hadamard matrices have a broad spectrum of applications:
\emph{Hadamard designs} \cite{beth:85} can be used for statistical experiments \cite{ragh:71}
and also for designing certain types of bridge tournaments \cite{berlekamp-hwang:72}; 
\emph{Hadamard codes} \cite{macw:98} were used in 1969 by space probe 
Mariner to transmit pictures from Mars to Earth \cite{posner:68}; 
and \emph{Hadamard transforms} \cite{nielsen:00} play an essential role in quantum computing, 
to name just a few. As a consequence natural connections arise between
the topics mentioned above and the results presented in this paper, and vice versa,
but we will not elaborate on them here.

\section{General properties of Bell inequalities} \label{S:prop_bell}
We will now use certain properties of matrices \eqref{E:hadamard} 
to derive some general properties of Bell inequalities. First, we note that all 
coefficients $a_k$ in \eqref{E:bell} are restricted to the values 
$0, \pm 2, \pm 4, \ldots, \pm 2^N$, which means, in particular, that they are even. 
Therefore, we can always divide \eqref{E:bell} by $2$ and still get an inequality with 
integral coefficients and an integral upper bound. Thus, any Bell inequality \eqref{E:bell}
can be written in the form
\begin{equation} \label{E:bell_gen}
\biggl\lvert\sum_{k=0}^{2^N-1} b_k E(k) \biggr\rvert \leq 2^{n}
\end{equation}
with integral $b_k$'s and $0 \leq n \leq N-1$. If the $b_k$'s are relatively prime and 
\begin{equation} \label{E:sum_cond}
\sum_{k=0}^{2^N-1} b_k > 0,
\end{equation}
we say that \eqref{E:bell_gen} is in \emph{standard form}.
(The sum \eqref{E:sum_cond} can never vanish, as we will see below; thus a division of
the $b_k$'s by their greatest common divisor and an occasional 
multiplication by $-1$ do the job.)

We will now derive some general properties of \eqref{E:bell_gen}, regardless of
whether it is in standard form or not:
\begin{lem} \label{L:bell_prop}
The coefficients in \eqref{E:bell_gen} have the following properties:
\begin{enumerate}[\upshape (i)]
\item For each $k$, there is $\lvert b_k \rvert \leq 2^n$. \label{I:bell_prop1}
\item If equality holds in \eqref{I:bell_prop1} for some $k$, then $b_j = 0$ for all $j \neq k$. \label{I:bell_prop2}
\item There is always \label{I:bell_prop3}
\begin{equation*}
\biggl\lvert\sum_{k=0}^{2^N-1} b_k \biggr\rvert = 2^n.
\end{equation*} 
\end{enumerate}
\end{lem}

\noindent\textit{Proof.}
We have already noted that the coefficients $a_k$ in \eqref{E:bell} are restricted to
the values $0, \pm 2, \pm 4, \ldots, \pm 2^N$. Hereby, the extreme values $\pm 2^N$ are taken 
exactly if the vector $(c_0,\ldots,c_{2^N-1})$ in \eqref{E:matrix} corresponds either to the $k$-th
row vector of $H_{2^N}$, or to this vector multiplied by $-1$. In both cases, 
we also have $a_j=0$ for $j \neq k$, since the row vectors of $H_{2^N}$ are orthogonal. 
Division of \eqref{E:bell} by $2^{N-n}$ leads to \eqref{E:bell_gen}, for which now 
properties \eqref{I:bell_prop1} and \eqref{I:bell_prop2} must hold. 

To show \eqref{I:bell_prop3}, we denote the column sums of $H_{2^N}$ 
by $\xi_k$ for $0 \leq k \leq 2^N-1$. By definition of $H_{2^N}$ we get
\begin{align*}
\xi_0 &= 2^N\\
\qquad \xi_k &= 0 \quad\text{for $k > 0$},
\end{align*}
and thus
\begin{equation} \label{E:sum_col}
\sum_{k=0}^{2^N-1} a_k = \sum_{k=0}^{2^N-1} \xi_k c_k = 2^N c_0.
\end{equation}
Taking absolute values in \eqref{E:sum_col} and dividing by $2^{N-n}$ 
leads to \eqref{I:bell_prop3}, which finishes the proof.\hfill\ensuremath{\Box}\medskip

We can easily verify property \eqref{I:bell_prop3} for the 
CHSH inequality \eqref{E:chsh2}. This property now also justifies the shorthand notation
\begin{equation*}
(b_0, b_1, \ldots, b_{2^N-1})
\end{equation*}
for \eqref{E:bell_gen}, since by its use we can always determine the upper bound in the corresponding 
Bell inequality.

The recursive principle in definition \eqref{E:hadamard} enables us to
take two arbitrary Bell inequalities for $N$ qubits and, by using them, construct 
a new Bell inequality for $N+1$ qubits. For that purpose we define the operator 
$\bowtie$ as
\begin{multline*}
\hspace*{-0.5em}(a_0,\ldots,a_{2^N-1}) \bowtie (b_0,\ldots,b_{2^N-1}) := \\
\quad(a_0+b_0,\ldots,a_{2^N-1}+b_{2^N-1},a_0-b_0,\ldots, a_{2^N-1}-b_{2^N-1}).
\end{multline*}
Thus in the first half of the resulting vector the coefficients are pairwisely added,
and in its second half they are pairwisely subtracted. By that we get:
\begin{thm} \label{T:bowtie}
Let $(a_0,\ldots,a_{2^N-1})$ and $(b_0,\ldots,b_{2^N-1})$ be two Bell inequalities for
$N$ qubits that satisfy
\begin{equation} \label{E:bound_cond}
\biggl\lvert\sum_{k=0}^{2^N-1} a_k\biggr\rvert = \biggl\lvert\sum_{k=0}^{2^N-1} b_k\biggr\rvert.
\end{equation}
Then applying the operator above
\begin{equation*}
(c_0,\ldots,c_{2^{N+1}-1}) := (a_0,\ldots,a_{2^N-1}) \bowtie (b_0,\ldots,b_{2^N-1})
\end{equation*}
yields a Bell inequality for $N+1$ qubits. By substituting all
Bell inequalities for $N$ qubits in $(a_0,\ldots,a_{2^N-1})$ and $(b_0,\ldots,b_{2^N-1})$,
we get all Bell inequalities for $N+1$ qubits.
\end{thm}

\noindent\textit{Proof.}
If we substitute $N \mapsto N+1$ in \eqref{E:matrix} and consider the form of the matrix
$H_{2^{N+1}}$ by setting $n=2^N$ in \eqref{E:hadamard}, we get exactly the transition
described by the $\bowtie$ operator. Therefore the result follows.\hfill\ensuremath{\Box}\medskip

By \eqref{E:matrix} we can also show that the rate of Bell inequalities that do not 
contain a certain expectation value decreases as $N$ increases. The following proposition
quantifies this fact:

\medskip\noindent\textbf{Proposition} {\itshape
The probability that an arbitrarily chosen Bell inequality for $N$ qubits does not contain a certain 
expectation value is asymptotically
\begin{equation*}
\frac{1}{\sqrt{2^{N-1}\pi}}
\end{equation*}
as $N \rightarrow \infty$.
}

\medskip\noindent\textit{Proof.}
Each coefficient in \eqref{E:bell} is by \eqref{E:matrix} the scalar product of a 
certain row vector of $H_{2^N}$ and a $\pm 1$-valued vector $(c_0,\ldots,c_{2^N-1})$. 
This product vanishes exactly if these two vectors differ in half of their elements. 
Since the number of vectors $(c_0,\ldots,c_{2^N-1})$ with this property is
\begin{equation} \label{E:binom}
\binom{2^N}{2^{N-1}},
\end{equation}
division by $2^{2^N}$ (the number of all inequalities) 
and Stirling's formula yield the result.\hfill\ensuremath{\Box}

\section{Bell polynomials} \label{S:poly}
It turns out to be convenient to assign to \eqref{E:bell_gen} 
the \emph{Bell polynomial}
\begin{equation} \label{E:poly}
B(z) := \sum_{k = 0}^{2^{N}-1} b_k z^k.
\end{equation}
(These polynomials should not be confused 
with the multivariate polynomials introduced by \emph{Eric Temple} Bell in 1934, or the 
polynomials defined in \cite{werner-wolf:01}.) If we want to emphasize the number of qubits for 
which \eqref{E:poly} is used, we will write $B^{(N)}(z)$ instead of $B(z)$. 

Property \eqref{I:bell_prop3} in Lemma~\ref{L:bell_prop}
tells us that the upper bound in the corresponding Bell inequality \eqref{E:bell_gen} 
is given by $\lvert B(1) \rvert$; by property \eqref{I:bell_prop1} this is
also an upper bound for the \emph{height} of that polynomial (which is defined as its maximal
coefficient, in absolute values). Multiplication of 
$B(z)$ by an arbitrary nonzero constant does not affect the corresponding
Bell inequality. Therefore we call polynomials that can be obtained from one another
by such a multiplication \emph{equivalent}. This definition indeed leads to an equivalence relation 
on the set of Bell polynomials. -- Usually we will choose representatives where the corresponding Bell 
inequality is in standard form, which means that the coefficients are relatively prime and $B(1) > 0$. 
(If we do not mind yielding rational coefficients, we can also consider polynomials with $B(1) = 1$ 
for that purpose.)

Having taken these preparations, we are now ready to adapt Lemma~\ref{L:bell_prop} 
and Theorem~\ref{T:bowtie} to Bell polynomials.
\begin{lem} 
The coefficients in \eqref{E:poly} have the following properties:
\label{L:poly_coeff}
\begin{enumerate}[\upshape (i)]
\item For each $k$, there is $\lvert b_k \rvert \leq \lvert B(1) \rvert$. \label{I:poly_coeff1}
\item If equality holds in \eqref{I:poly_coeff1} for some $k$, then $b_j = 0$ for all $j \neq k$.
\end{enumerate}
\end{lem}

\noindent\textit{Proof.}
The lemma follows immediately from the definition of Bell polynomials and 
Lemma~\ref{L:bell_prop}.\hfill\ensuremath{\Box}\medskip

\noindent For the adaptation of Theorem~\ref{T:bowtie} we observe that
\begin{equation*}
\begin{array}{llllrrcr}
(a_0,&\ldots,& a_{k-1},&0,&\ldots,&0) &\ \leftrightarrow\ &  A(z)\phantom{.}\\
(0,&\ldots,&0,&a_0,&\ldots,& a_{k-1}) &\ \leftrightarrow\ &  z^k A(z).
\end{array}
\end{equation*}
This leads us to define
\begin{equation*}
A(z) \bowtie B(z) := \bigl(1+z^{2^N}\bigr)A(z) + \bigl(1-z^{2^N}\bigr)B(z),
\end{equation*}
where $A(z)$ and $B(z)$ are Bell polynomials for $N$ qubits. By that we get:
\begin{thm} \label{T:bowtie_poly}
Let $A(z)$ and $B(z)$ be two Bell polynomials for $N$ qubits that satisfy
\begin{equation} \label{E:bound_cond_poly}
\bigl\lvert A(1) \bigr\rvert = \bigl\lvert B(1) \bigr\rvert.
\end{equation}
Then applying the operator above
\begin{equation*}
C(z) := A(z) \bowtie B(z)
\end{equation*}
yields a Bell polynomial for $N+1$ qubits. By substituting all Bell polynomials for 
$N$ qubits in $A(z)$ and $B(z)$, we get all Bell polynomials for $N+1$ qubits.
\end{thm}

\noindent\textit{Proof.}
The theorem follows immediately from the definition of Bell polynomials and 
Theorem~\ref{T:bowtie}.\hfill\ensuremath{\Box}\medskip

\noindent\textbf{Example}
Let us consider the two CHSH inequalities
\begin{equation} \label{E:chsh_ex}
(1,1,1,-1) \quad\text{and}\quad (1,-1,-1,-1).
\end{equation}
They obviously satisfy \eqref{E:bound_cond}, hence we can apply Theorem~\ref{T:bowtie}
and get
\begin{equation*}
(1,1,1,-1) \bowtie (1,-1,-1,-1) = (2,0,0,-2,0,2,2,0).
\end{equation*}
After division by $2$, in traditional notation this reads as
\begin{equation*}
\lvert E(1,1,1)-E(1,2,2)+E(2,1,2)+E(2,2,1) \rvert \leq 2,
\end{equation*}
which is an MABK inequality \cite{mermin:90a, arde:92, belin:93}. Alternatively, we can use the 
corresponding Bell polynomials and Theorem~\ref{T:bowtie_poly} to achieve the same result. 
In that case we have
\begin{alignat*}{4}
(1,1,1,-1) &\ \leftrightarrow\ &  A(z) &= 1+z+z^2-z^3\\
(1,-1,-1,-1) &\ \leftrightarrow\ & B(z) &= 1-z-z^2-z^3
\end{alignat*}
and 
\begin{alignat*}{2}
A(z) \bowtie B(z) =&\, (1+z^4)(1+z+z^2-z^3)+\\
		       &\, (1-z^4)(1-z-z^2-z^3)\\
                  =&\, 2-2z^3+2z^5+2z^6,
\end{alignat*}
which indeed corresponds to the inequality derived above. 
(Note also that whenever \eqref{E:bound_cond} is satisfied, the same is automatically 
true for the corresponding Bell polynomials and \eqref{E:bound_cond_poly}.)

If we replace the second CHSH inequality in \eqref{E:chsh_ex} by the
trivial inequality $(1,0,0,0)$, we must first multiply this inequality by $2$
in order to satisfy condition \eqref{E:bound_cond}. Application of Theorem~\ref{T:bowtie}
now yields
\begin{equation*}
(1,1,1,-1) \bowtie (2,0,0,0) = (3,1,1,-1,-1,1,1,-1),
\end{equation*}
which corresponds to
\begin{equation*}
3+z+z^2-z^3-z^4+z^5+z^6-z^7.
\end{equation*}
(This time we leave it to the reader to obtain that polynomial directly by use of 
Theorem~\ref{T:bowtie_poly}.) In traditional notation, this reads as
\begin{multline*}
\lvert 3 E(1,1,1) + E(1,1,2) + E(1,2,1) - E(1,2,2) \\
      \quad - E(2,1,1) + E(2,1,2) + E(2,2,1) - E(2,2,2) \rvert \leq 4,
\end{multline*}
which is already in standard form.\hfill\ensuremath{\Diamond}\medskip

We will now study the structure of Bell polynomials in detail.
For $N=1$, the whole set of polynomials is given by $\pm 1$ and $\pm z$. 
Theorem~\ref{T:bowtie_poly} tells us that the Bell polynomials for $N=2$ have the form
\vspace*{-.35cm}
\begin{equation*}
\begin{array}{rrllrl}
           &           & \raisebox{-.2cm}{\!\!$\pm 1$} &             &            & \raisebox{-.2cm}{\!\!$\pm 1$} \\
           & \raisebox{-.05cm}{\!\!\!$\diagup$}   &   &             & \raisebox{-.05cm}{\!\!\!$\diagup$}    &   \\
(1+z^2) &           &   & \!\!+\ \ (1-z^2) &            &   \\
           & \raisebox{.05cm}{\!\!\!$\diagdown$} &   &             & \raisebox{.05cm}{\!\!\!$\diagdown$} &   \\
           &           & \raisebox{.2cm}{\!\!$\pm z$} &             &            & \raisebox{.2cm}{\!\!$\pm z$}
\end{array}
\vspace*{-.35cm}
\end{equation*}
(presented in a hopefully self-explanatory notation).
Proceeding in that way, we see that $B^{(N)}(z)$ consists of the $2^{N-1}$ summands
\begin{equation} \label{E:summand}
\bigl(1 \pm z^{2^{N-1}}\bigr) \bigl(1 \pm z^{2^{N-2}}\bigr) \cdots \bigl(1 \pm z^{2}\bigr)
\end{equation}
with each of them multiplied by a factor $\pm 1$ or $\pm z$. Since there are exactly four choices 
for this multiplication within each summand, the number of Bell polynomials is indeed $4^{2^{N-1}} = 2^{2^N}$.

It turns out to be convenient to choose a fixed enumeration for the summands \eqref{E:summand}. 
Therefore, we define
\begin{equation} \label{E:s_k}
s_k^{(N)}(z) := \bigl(1 + (-1)^{k_{N-2}} z^{2^{N-1}}\bigr) \cdots \bigl(1 + (-1)^{k_0} z^{2}\bigr)
\end{equation}
with
\begin{equation*}
k = (k_{N-2},\ldots,k_1,k_0)_2 = \sum_{i=0}^{N-2} k_i 2^i,
\end{equation*}
whereby the empty product is set $1$ as usual.
Thus, we have
\begin{gather*}
\begin{aligned}
s_0^{(1)}(z) &= 1\\[.15cm]
s_0^{(2)}(z) &= 1+z^2\\
s_1^{(2)}(z) &= 1-z^2\\[.15cm]
s_0^{(3)}(z) &= (1+z^4)(1+z^2)\\
s_1^{(3)}(z) &= (1+z^4)(1-z^2)\\
s_2^{(3)}(z) &= (1-z^4)(1+z^2)\\
s_3^{(3)}(z) &= (1-z^4)(1-z^2)\\[.15cm]
s_0^{(4)}(z) &= (1+z^8)(1+z^4)(1+z^2)\\
s_1^{(4)}(z) &= (1+z^8)(1+z^4)(1-z^2)
\end{aligned}\\
\cdots\cdots\cdots\cdots\cdots\cdots\cdots\cdots\cdots\cdots
\end{gather*}
We can also use a recursive definition for $s_k^{(N)}(z)$, namely
\begin{align} 
s_0^{(1)}(z) &:= 1 \notag \\
s_k^{(N)}(z) &:= \begin{cases} \bigl(1+z^{2^{N-1}}\bigr) s_k^{(N-1)}(z)           & \text{if $k < 2^{N-2}$}\\[.15cm]
                               \bigl(1-z^{2^{N-1}}\bigr) s_{k-2^{N-2}}^{(N-1)}(z) & \text{if $k \geq 2^{N-2}$}
                \end{cases} \label{E:summand_rec}
\end{align}
with $0 \leq k < 2^{N-1}$. Expanding $s_k^{(N)}(z)$ yields an expression of the form
\begin{equation} \label{E:even}
1 \pm z^2 \pm z^4 \pm \cdots \pm z^{2^N-2},
\end{equation}
thus $s_k^{(N)}(z)$ is an even polynomial with coefficients $\pm 1$ and degree $2^N-2$.
Multiplication by $z$ yields 
\begin{equation} \label{E:odd}
z \pm z^3 \pm z^5 \pm \cdots \pm z^{2^N-1},
\end{equation}
which now is an odd polynomial with coefficients $\pm 1$ and degree $2^N-1$.

\begin{thm} \label{T:main}
Let $s_k^{(N)}(z)$ be as defined in \eqref{E:summand_rec}. Then the complete set of Bell polynomials 
for $N$ qubits is given by
\begin{equation} \label{E:bell_main}
B_{uv}^{(N)}(z) = \sum_{k=0}^{2^{N-1}-1} (-1)^{u_k} z^{v_k} s_k^{(N)}(z),
\end{equation}
where $u = (u_{2^{N-1}-1},\ldots, u_0)_2$ and $v = (v_{2^{N-1}-1},\ldots, v_0)_2$ 
are arbitrary numbers with binary expansions of length $2^{N-1}$ (occasionally
written with leading zeros).
\end{thm}

\noindent\textit{Proof.}
We have already achieved the structure of these polynomials in the recursive process 
that led to the definition of $s_k^{(N)}(z)$. The different choices of factors $\pm 1$ and $\pm z$ 
at the $k$-th summand are now described by the Boolean variables $u_k$ and $v_k$, 
so we are done.\hfill\ensuremath{\Box}\medskip

\noindent\textit{Remark.}
A more direct approach would be to define the polynomials
\begin{equation} \label{E:t_k}
t_k^{(N)}(z):= h_{0k} + h_{1k}z + h_{2k}z^2 + \cdots + h_{N-1,k}z^{N-1}
\end{equation}
by using \eqref{E:h_jk}, and consider
\begin{equation*}
B_{w}^{(N)}(z) := \frac{1}{2}\sum_{k=0}^{2^N-1} (-1)^{w_k} t_k^{(N)}(z)
\end{equation*}
for an arbitrary $w = (w_{2^N-1},\ldots, w_0)_2$. However, the form \eqref{E:bell_main} reveals more 
of the structure of these polynomials, which turns out to be useful. 
The connection between \eqref{E:s_k} and \eqref{E:t_k} is given by
\begin{equation} \label{E:s_t_conn}
s_k^{(N)}(z) = t_k^{(N-1)}(z^2)
\end{equation}
for $0 \leq k < 2^{N-1}$.

\medskip\noindent\textbf{Example}
For $N=1$ and $N=2$ the complete set of Bell polynomials is given by
\begin{equation*}
\begin{array}{l@{\ }c@{\ }l@{\qquad}l@{\ }c@{\ }l}
B_{00}^{(1)}(z) &=& 1 & B_{10}^{(1)} &=& -1\\
B_{01}^{(1)}(z) &=& z & B_{11}^{(1)} &=& -z
\end{array}
\end{equation*}
and
\begin{equation*}
\begin{array}{l@{\ }c@{\ }l@{\quad\!}l@{\ }c@{\ }l}
\!\!B_{00}^{(2)}(z) &=& \phantom{-}2            & B_{20}^{(2)}(z) &=& \phantom{-}2z^2\\
\!\!B_{01}^{(2)}(z) &=& \phantom{-}1+z-z^2+z^3  & B_{21}^{(2)}(z) &=& -1+z+z^2+z^3\\
\!\!B_{02}^{(2)}(z) &=& \phantom{-}1+z+z^2-z^3  & B_{22}^{(2)}(z) &=& \phantom{-}1-z+z^2+z^3\\
\!\!B_{03}^{(2)}(z) &=& \phantom{-}2z           & B_{23}^{(2)}(z) &=& \phantom{-}2z^3\\[.25cm]
\!\!B_{10}^{(2)}(z) &=& -2z^2                   & B_{30}^{(2)}(z) &=& -2\\
\!\!B_{11}^{(2)}(z) &=& \phantom{-}1-z-z^2-z^3  & B_{31}^{(2)}(z) &=& -1-z+z^2-z^3\\
\!\!B_{12}^{(2)}(z) &=& -1+z-z^2-z^3            & B_{32}^{(2)}(z) &=& -1-z-z^2+z^3\\
\!\!B_{13}^{(2)}(z) &=& -2z^3                   & B_{33}^{(2)}(z) &=& -2z.
\end{array}
\end{equation*}
Note again that all nontrivial cases above correspond to CHSH inequalities.\hfill\ensuremath{\Diamond}

\begin{lem}
Let $B_{uv}^{(N)}(z)$ be as defined in \eqref{E:bell_main}. Then
\begin{alignat}{4}
&B_{uv}^{(N)}(1)  &&= (-1)^{u_0}\,2^{N-1} \label{E:arg1}\\
&B_{uv}^{(N)}(-1) &&= (-1)^{u_0+v_0}\,2^{N-1} \label{E:arg-1}\\
&B_{uv}^{(N)}(0)  &&= \sum_{k=0}^{2^{N-1}-1} (-1)^{u_k}(1-v_k). \label{E:arg0}
\end{alignat}
Furthermore we have
\begin{align}
-B_{uv}^{(N)}(z) &= B_{\hat{u}v}^{(N)}(z) \label{E:poly_trans1} \\
B_{uv}^{(N)}(-z) &= B_{u \oplus v,v}^{(N)}(z), \label{E:poly_trans2}
\end{align}
where $\hat{u}$ and $u \oplus v$ are defined as
\begin{align}
\hat{u} &:= (1-u_{2^{N-1}-1},\ldots, 1-u_0)_2 \label{E:u-hat} \\
u \oplus v &:= (u_{2^{N-1}-1} \oplus v_{2^{N-1}-1},\ldots, u_0 \oplus v_0)_2 \label{E:u-xor-v}
\end{align}
and the operator $\oplus$ on the right-hand side of \eqref{E:u-xor-v} means addition over 
$\text{\rm GF}(2)$. (Thus, $u \oplus v$ defines a bitwise ``exclusive or'' operation on 
$u$ and $v$.)
\end{lem}

\noindent\textit{Proof.}
For all $k>0$ there is 
\begin{equation} \label{E:arg1-1}
s_k^{(N)}(1) = s_k^{(N)}(-1) = 0,
\end{equation}
since in that case $s_k^{(N)}(z)$ always contains a factor $1-z^{2^j}$ for some $j$. 
Therefore, by \eqref{E:bell_main} and 
\begin{equation*}
s_0^{(N)}(1)=2^{N-1}
\end{equation*} 
we immediately get \eqref{E:arg1} and \eqref{E:arg-1}. 
Similarly, for any $N$ and $k$ there is 
\begin{equation*}
s_k^{(N)}(0)=1.
\end{equation*}
Thus, by setting $z=0$ in \eqref{E:bell_main} we get the following: The $k$-th summand
contributes only to the sum if $v_k=0$, namely $+1$ if $u_k=0$ and $-1$ if $u_k=1$. 
This is exactly what \eqref{E:arg0} tells us formally. 

The transformation $B_{uv}^{(N)}(z) \mapsto -B_{uv}^{(N)}(z)$ can be achieved by 
substituting $u_k \mapsto 1-u_k$ for $0 \leq k < 2^{N-1}$, since the latter changes the
sign of every summand in \eqref{E:bell_main}; therefore \eqref{E:poly_trans1} holds. 
Finally, the transformation $z \mapsto -z$ changes the sign of the $k$-th summand
exactly if $v_k=1$, since $s_k^{(N)}(z)$ is even for any $N$ and $k$. 
This shows that \eqref{E:poly_trans2} holds, and thus completes the 
proof.\hfill\ensuremath{\Box}\medskip

\begin{figure*}%
\includegraphics{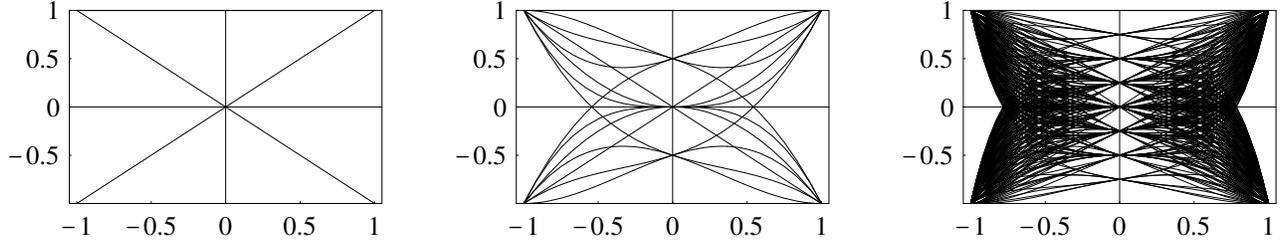}
\caption{The normalized Bell polynomials $\widetilde{B}_{uv}^{(N)}(z)$ for $1 \leq N \leq 3$.}
\label{F:poly}
\end{figure*}

\noindent\textbf{Corollary} {\itshape
The polynomials \eqref{E:bell_main} satisfy
\begin{equation} \label{E:abs_one}
\bigl\lvert B_{uv}^{(N)}(1) \bigr\rvert = \bigl\lvert B_{uv}^{(N)}(-1) \bigr\rvert = 2^{N-1}.
\end{equation}
Furthermore, $B_{uv}^{(N)}(z)$ is even exactly if $v=0$, and odd 
exactly if $v=2^{2^{N-1}}-1$.
}

\medskip\noindent\textit{Proof.}
Property \eqref{E:abs_one} is a trivial consequence of \eqref{E:arg1} and \eqref{E:arg-1}.
If $B_{uv}^{(N)}(z)$ is even, then by \eqref{E:poly_trans2} we get
\begin{equation*}
B_{uv}^{(N)}(z) = B_{u \oplus v,v}^{(N)}(z).
\end{equation*}
Hence $u = u \oplus v$, which means that
\begin{equation*}
v=(0,0,\ldots, 0)_2=0.
\end{equation*}
Conversely, if $v=0$, then $B_{uv}^{(N)}(z) = B_{uv}^{(N)}(-z)$ by \eqref{E:poly_trans2},
and $B_{uv}^{(N)}(z)$ is thus even. 

On the other hand, if $B_{uv}^{(N)}(z)$ is odd, then by \eqref{E:poly_trans1} and
\eqref{E:poly_trans2} we get
\begin{equation*}
B_{\hat{u}v}^{(N)}(z) = B_{u \oplus v,v}^{(N)}(z).
\end{equation*}
Hence $\hat{u} = u \oplus v$, which means that
\begin{equation*}
v=(1,1,\ldots, 1)_2 = 2^{2^{N-1}}-1.
\end{equation*}
Conversely, if $v=2^{2^{N-1}}-1$, then $B_{uv}^{(N)}(-z) = -B_{uv}^{(N)}(z)$
by \eqref{E:poly_trans1} and \eqref{E:poly_trans2}, and $B_{uv}^{(N)}(z)$ is thus odd. 
So we are done.\hfill\ensuremath{\Box}\medskip

Because of \eqref{E:abs_one}, it is sometimes useful to consider the \emph{normalized} Bell polynomials 
\begin{equation} \label{E:poly_norm}
\widetilde{B}_{uv}^{(N)}(z) := 2^{1-N} B_{uv}^{(N)}(z),
\end{equation}
which have the property that
\begin{equation*}
\bigl\lvert\widetilde{B}_{uv}^{(N)}(1)\bigr\rvert = 
\bigl\lvert\widetilde{B}_{uv}^{(N)}(-1)\bigr\rvert = 1.
\end{equation*} 
For $1 \leq N \leq 3$, they are depicted in Fig.~\ref{F:poly}.

\section{Analyzing the structure of Bell inequalities}
The numbers $u$ and $v$ completely determine the polynomial \eqref{E:bell_main} for any fixed 
$N$. Since the bits in the binary expansion of $u$ coincide with the signs of the
summands in \eqref{E:bell_main}, we call $u$ the \emph{sign number} of $B_{uv}^{(N)}(z)$.
On the other hand, the bits in the binary expansion of $v$ describe whether 
a summand in \eqref{E:bell_main} is an even or an odd polynomial (of form \eqref{E:even} or 
\eqref{E:odd}, respectively). Therefore we call $v$ the \emph{parity number} of 
$B_{uv}^{(N)}(z)$.

Since $B_{uv}^{(N)}(z)$ and $B_{\hat{u}v}^{(N)}(z)$ are equivalent by \eqref{E:poly_trans1}, 
it is sufficient to consider even sign numbers. (This is particularly the case if the
corresponding Bell inequality is in standard form, since from $B_{uv}^{(N)}(1) > 0$ and \eqref{E:arg1}
it follows that $u_0=0$.) Furthermore, we can extend the notion
of equivalence to enclose \emph{symmetry transformations} as well. This means that we do not distinguish 
between inequalities that can be transformed into one another by permutations of sites, observables  
or measurement values (since those can be considered as not ``essentially different'').
This clearly reduces the number of equivalence classes for any fixed $N$,
though it does not substantially affect the growth rate of this number as $N \rightarrow \infty$.

We will now study properties of Bell polynomials in terms of $u$ and $v$.
But before we do this we make some simple observations:
\begin{lem} \label{L:s}
Let $s_k^{(N)}(z)$ be as defined in \eqref{E:summand_rec}. Then:
\begin{enumerate}[\upshape (i)]
\item All coefficients in $s_0^{(N)}(z)$ are $+1$. \label{I:s1}
\item For $k>0$ half of the coefficients in $s_k^{(N)}(z)$ are $+1$, the other half being $-1$. \label{I:s2}
\item For each $N$, there is \label{I:s3}
\begin{equation*}
\sum_{k=0}^{2^{N-1}-1} s_k^{(N)}(z) = 2^{N-1}.
\end{equation*}
\end{enumerate}
\end{lem}

\noindent\textit{Proof.}
By using \eqref{E:s_t_conn}, the content of this lemma is an immediate consequence of the  
properties of the Hadamard matrices investigated in Section~\ref{S:prop_bell}. 
We can also prove this lemma directly:
Property \eqref{I:s1} is obvious since $s_0^{(N)}(z)$ consists only of 
factors $1+z^{2^j}$, which all have positive signs. By \eqref{E:arg1-1} we have already seen 
that $s_k^{(N)}(1) = 0$ for $k>0$. If, on the other hand, we set $z=1$ in \eqref{E:even},
we get an expression of the form 
\begin{equation*}
1 \pm 1 \pm 1 \pm \cdots \pm 1.
\end{equation*}
This sum can only be zero if the number of $+1$'s equals the number of $-1$'s,
thus \eqref{I:s2} holds. Finally, \eqref{I:s3} can be proved by induction 
(which would have also been possible in case of \eqref{I:s2}). 
Obviously \eqref{I:s3} holds for $N=1$, so let us assume by induction 
hypothesis that it is true for $N-1$. By setting
\begin{equation*}
S_N := \sum_{k=0}^{2^{N-1}-1} s_k^{(N)}(z),
\end{equation*}
this means that $S_{N-1}=2^{N-2}$. Now by \eqref{E:summand_rec} we get
\begin{align*}
S_N &= \sum_{k=0}^{2^{N-2}-1} s_k^{(N)}(z) + \sum_{k=2^{N-2}}^{2^{N-1}-1} s_k^{(N)}(z)\\
                                    &= \bigl(1+z^{2^{N-1}}\bigr)S_{N-1} + \bigl(1-z^{2^{N-1}}\bigr)S_{N-1}\\
                                    &= 2 S_{N-1} = 2^{N-1}.
\end{align*}
This concludes the proof.\hfill\ensuremath{\Box}\medskip

\noindent\textit{Hint.} 
Property \eqref{I:s3} can be generalized for arbitrary sums of terms 
\begin{equation*}
\bigl(1 \pm z^{k_1}\bigr)\bigl(1 \pm z^{k_2}\bigr) \cdots \bigl(1 \pm z^{k_n}\bigr)
\end{equation*}
over all possible sign combinations. By symmetry we see that this sum is always $2^n$, 
independently of the concrete values of the $k_i$'s. We can therefore immediately 
simplify an expression like
\begin{equation*}
(1+z^8)(1+z^2)+(1+z^8)(1-z^2)+(1-z^8)(1-z^2)
\end{equation*}
to
\begin{equation*}
4-(1-z^8)(1+z^2) = 3-z^2+z^8+z^{10},
\end{equation*}
since the sum over all terms $(1 \pm z^8)(1 \pm z^2)$ must be~$4$. The sketched method can frequently 
be applied in the process of determining Bell polynomials 
(see the example below).
\begin{thm} \label{T:poly_prop}
For any Bell polynomial 
\begin{equation} \label{E:bell_poly}
B_{uv}(z) = \sum_{k = 0}^{2^N-1} b_k z^k,
\end{equation}
as defined in \eqref{E:bell_main}, we have:
\begin{enumerate}[\upshape (i)]
\item If $u=0$, then \label{I:poly_prop1}
\begin{align*}
b_0 &= 2^{N-1} - b_1\\
b_2 &= - b_3\\
b_4 &= - b_5\\
&\hspace*{-.2cm}\cdots\cdots\\
b_{2^N-2} &= - b_{2^N-1}.
\end{align*} 
\item If $v=0$, then $b_{2j+1} = 0$ for all $j \geq 0$. \label{I:poly_prop2}
\item If $v$ is even, then \label{I:poly_prop3}
\begin{equation*}
b_1 + b_3 + \cdots + b_{2^N-1} = 0.
\end{equation*}
\item If $v$ is odd, then \label{I:poly_prop4}
\begin{equation*}
b_0 + b_2 + \cdots + b_{2^N-2} = 0.
\end{equation*}
\item If the binary expansion of $v$ contains an even (odd) number of $1$'s, 
all $b_k$'s are even (odd). \label{I:poly_prop5}
\end{enumerate}
\end{thm}

\noindent\textit{Proof.}
Let
\begin{align}
p(z) &:= \sum_{k = 0}^{2^{N-1}-1} (-1)^{u_k}(1-v_k) s_k^{(N)}(z) \label{E:p} \\
q(z) &:= \sum_{k = 0}^{2^{N-1}-1} (-1)^{u_k} v_k s_k^{(N)}(z),   \label{E:q}
\end{align}
then
\begin{equation*}
B_{uv}(z) = p(z) + z\mspace{1mu}q(z).
\end{equation*}
In other words, $p(z)$ denotes the even part of $B_{uv}(z)$, and $z\mspace{1mu}q(z)$ denotes 
its odd part (because $p(z)$ and $q(z)$ are both even). If $u=0$, then by property~\eqref{I:s3} 
in Lemma~\ref{L:s} we have
\begin{equation*}
p(z)=2^{N-1}-q(z),
\end{equation*}
therefore \eqref{I:poly_prop1} holds. On the other hand, if $v=0$ then $q(z) \equiv 0$, and
now \eqref{I:poly_prop2} follows (since in that case the polynomial $B_{uv}(z)$ is even). 

To show \eqref{I:poly_prop3} and \eqref{I:poly_prop4} we use \eqref{E:arg1} and \eqref{E:arg-1},
by which we get
\begin{alignat*}{3}
&b_0 + b_1 + b_2 + &\cdots + b_{2^N-2} + b_{2^N-1} &= (-1)^{u_0}2^{N-1}      \\
&b_0 - b_1 + b_2 - &\cdots + b_{2^N-2} - b_{2^N-1} &= (-1)^{u_0+v_0}2^{N-1}.
\end{alignat*}
Subtraction and addition of these equalities now prove \eqref{I:poly_prop3} and 
\eqref{I:poly_prop4}, respectively (since $v_0=0$ if $v$ is even, and $v_0=1$ otherwise). 
Finally we get \eqref{I:poly_prop5} by observing that an even (odd) number of nonvanishing 
summands in \eqref{E:p} and \eqref{E:q} also leads to even (odd) coefficients in 
these polynomials.\hfill\ensuremath{\Box}\medskip

\noindent\textit{Remark.}
The theorem above can be enhanced in the following ways:
\begin{enumerate}[\upshape (a)]
\item Properties \eqref{I:poly_prop2}--\eqref{I:poly_prop4} are also valid 
for equivalent polynomials of $B_{uv}(z)$, whereas properties \eqref{I:poly_prop1} and 
\eqref{I:poly_prop5} need some slight adaptation if equivalent polynomials of $B_{uv}(z)$ are 
considered. 
\item Similar results to
\eqref{I:poly_prop1} and \eqref{I:poly_prop2} can be derived for other specific values
of $u$ and $v$, such as 
\begin{equation*}
u=2^{2^{N-1}}-1 \quad \text{or} \quad v=2^{2^{N-1}}-1.
\end{equation*}
\item If $B_{uv}^{(k)}(z)$ denotes (by a short-term overload of notation) 
the $k$-th derivative of $B_{uv}(z)$, then
\begin{equation*}
b_k = \frac{1}{k!} B_{uv}^{(k)}(0).
\end{equation*}
We can therefore formulate properties of the $b_k$'s also in terms of derivatives of 
$B_{uv}(z)$; e.g.\ property \eqref{I:poly_prop1} then reads as
\begin{align*}
B'_{0v}(0) &= 2^{N-1} - B_{0v}(0)\\
B_{0v}^{(2j+1)}(0) &= - (2j+1) B_{0v}^{(2j)}(0) \quad \text{for $j \geq 1$}.
\end{align*}
\item The $b_k$'s can also be written in terms of $u$ and $v$. 
For instance
\begin{equation} \label{E:b0}
b_0 = \langle \hat{u},\hat{v} \rangle - \langle u,\hat{v} \rangle,
\end{equation}
where $\langle u, v \rangle$ denotes the scalar product of the binary expansions of $u$ and $v$
(which counts the positions of $1$'s at which the binary expansions of $u$ and $v$ coincide).
\item By \eqref{E:b0} we can again count the Bell inequalities that do not contain $E(0)$, as was 
done for arbitrary expectation values in the proof of the proposition at the end
of Section~\ref{S:prop_bell}.
Now $b_0=0$ holds if and only if exactly half of the $1$'s in the binary expansion of 
$u$ coincide with $1$'s in the binary expansion of $v$; together with \eqref{E:binom}, 
this proves the formula
\begin{equation*}
\sum_{k=0}^{2^{N-2}}\binom{2^{N-1}}{2k}\binom{2k}{k}2^{2^{N-1}-2k} = \binom{2^N}{2^{N-1}}.
\end{equation*}
\end{enumerate}

\medskip\noindent\textbf{Example}
Let $N=4$ and
\begin{align*}
u &= (0,0,0,0,1,1,1,0)_2\\
v &= (0,0,0,0,0,0,0,0)_2.
\end{align*}
Then 
\begin{align*}
B_{uv}(z) &=(1+z^8)(1+z^4)(1+z^2)\\
          &\phantom{=} -(1+z^8)(4-(1+z^4)(1+z^2))\\
          &\phantom{=} +4(1-z^8)\\
          &=2(1+z^2+z^4+z^6-3z^8+z^{10}+z^{12}+z^{14}),
\end{align*}
using the hint provided after the proof of Lemma~\ref{L:s}.
We can now easily verify properties \eqref{I:poly_prop2} and \eqref{I:poly_prop5} of the theorem above.
(Actually, property \eqref{I:poly_prop3} also holds, but this is trivial if \eqref{I:poly_prop2} can 
be applied.) If we consider alternatively
\begin{align*}
u &= (0,0,0,0,0,0,0,0)_2\\
v &= (0,0,0,0,1,1,1,0)_2,
\end{align*}
we get
\begin{align*}
B_{uv}(z) &=(1+z^8)(1+z^4)(1+z^2)\\
          &\phantom{=} +z(1+z^8)(4-(1+z^4)(1+z^2))\\
          &\phantom{=} +4(1-z^8)\\
          &=5+3z+z^2-z^3+z^4-z^5+z^6-z^7\\
          &\phantom{=} -3z^8+3z^9+z^{10}-z^{11}+z^{12}-z^{13}+z^{14}-z^{15}.
\end{align*}
This time properties \eqref{I:poly_prop1}, \eqref{I:poly_prop3} and 
\eqref{I:poly_prop5} of the theorem above can be verified.\hfill\ensuremath{\Diamond}\medskip

From now on we will call a Bell inequality \emph{full-term} if it contains all expectation values  
under consideration. 
(We can further call a Bell inequality \emph{$t$-term} if it 
contains precisely $t$ expectation values; then the full-term inequalities for $N$ qubits are
$2^N$-term, and the trivial inequalities are exactly the $1$-term inequalities.)
Using that diction we get:
\begin{thm}
Any complete set of inequivalent Bell inequalities for $N$ qubits has the following characteristics: 
\begin{enumerate}[\upshape (A)]
\item Exactly $2^N$ inequalities are trivial. \label{I:term1}
\item At least half of the inequalities are full-term. \label{I:term2}
\end{enumerate}
\end{thm}

\noindent\textit{Proof.}
The number of expectation values that can possibly appear in a Bell inequality for $N$ qubits
is $2^N$. Up to equivalence, exactly one trivial Bell inequality corresponds to any of these values; 
therefore \eqref{I:term1} follows. 

By property~\eqref{I:poly_prop5} in Theorem~\ref{T:poly_prop}
we know that an odd number of $1$'s in the binary expansion of $v$ leads to odd coefficients
in \eqref{E:bell_poly}. In particular, these coefficients are therefore all different from zero.
Since half of all $v$'s have the property stated above, at least half of the corresponding Bell 
inequalities are full-term. This fact still holds if equivalent polynomials are eliminated from that
set, which proves \eqref{I:term2}.\hfill\ensuremath{\Box}\medskip

Actually, for $N=2$ and $N=3$ 
\emph{exactly} half of the Bell inequalities are full-term. Thus, in these cases there is 
a one-to-one correspondence to $v$'s with an odd number of $1$'s in 
their binary expansion. It follows that each of these inequalities, in standard form, 
has the upper bound $2^{N-1}$.

We give a concluding example in order to illustrate how the results above can be used to 
construct Bell inequalities with certain properties:

\medskip\noindent\textbf{Example}
Find all Bell inequalities for $N$ qubits in standard form, where the 
coefficient of $E(k,k,\ldots,k)$ is maximal.

\medskip\noindent\textit{Solution.} 
We have already listed the Bell polynomials for $N=1$ and $N=2$, so let us assume that $N \geq 3$.
We will first consider the case $k=0$. 
By \eqref{E:bell_gen} and property~\eqref{I:bell_prop1} in Lemma~\ref{L:bell_prop} we know 
that the upper bound in any such 
inequality, in standard form, is less than or equal $2^{N-1}$. Property~\eqref{I:bell_prop2}
in the same lemma tells us that the coefficient $b_0=2^{N-1}$ can hereby never be 
admitted, since in that case the inequality can always be reduced to
\begin{equation*}
\lvert E(0,0,\ldots,0) \rvert \leq 1.
\end{equation*}
Thus we are aimed to consider $b_0=2^{N-1}-1$. 

By property~\eqref{I:poly_prop5} in 
Theorem~\ref{T:poly_prop} all coefficients of the corresponding inequality are odd; 
otherwise, this inequality could again be reduced and property~\eqref{I:bell_prop1} in Lemma~\ref{L:bell_prop}
would be violated. Consequently, all inequalities of the desired type are full-term.
The former property tells us further that the binary expansion of $v$ contains an odd 
number of $1$'s. Together with \eqref{E:b0}, which reads as
\begin{equation*}
\langle \hat{u},\hat{v} \rangle - \langle u,\hat{v} \rangle = 2^{N-1}-1,
\end{equation*}
this leads to 
\begin{equation*}
\langle \hat{u},\hat{v} \rangle = 2^{N-1}-1 \quad \text{and} \quad \langle u,\hat{v} \rangle = 0.
\end{equation*}
Recalling the definition of standard form, the corresponding Bell polynomial $B(z)$ must 
also satisfy $B(1)>0$; thus, by \eqref{E:arg1} we further have $u_0=0$. 
All in all, that means the solutions for $k=0$ correspond to
\begin{gather*}
u=(0,0,\ldots,0,0,0)_2 \quad \text{and} \quad v=(0,0,\ldots,0,0,1)_2,\\
u=(0,0,\ldots,0,\ast,0)_2 \quad \text{and} \quad v=(0,0,\ldots,0,1,0)_2,\\
u=(0,0,\ldots,\ast,0,0)_2 \quad \text{and} \quad v=(0,0,\ldots,1,0,0)_2,\\
\cdots\cdots\cdots\cdots\cdots\cdots\cdots\cdots\cdots\cdots\cdots\cdots\cdots\cdots\cdots\cdots\\
u=(0,\ast,\ldots,0,0,0)_2 \quad \text{and} \quad v=(0,1,\ldots,0,0,0)_2,\\
u=(\ast,0,\ldots,0,0,0)_2 \quad \text{and} \quad v=(1,0,\ldots,0,0,0)_2,
\end{gather*}
where $\ast$ denotes an arbitrary bit. Hence the number of corresponding Bell polynomials is  
$2^N-1$. The polynomials for $k=1$ are finally reached by the symmetry transformation
\begin{equation*}
B(z) \mapsto z^{2^N-1}\,B\Bigl(\frac{1}{z}\Bigr),
\end{equation*}
since this is exactly what happens if the enumeration of observables is reversed
(and thus $k \mapsto 1-k$).

To see a concrete instance of polynomials, let us consider the case $N=3$. 
The desired Bell polynomials for $k=0$ are
\begin{align*}
&3+z-z^2+z^3-z^4+z^5-z^6+z^7 \\
&3+z+z^2-z^3-z^4+z^5+z^6-z^7 \\
&3-z+z^2+z^3-z^4-z^5+z^6+z^7 \\
&3+z-z^2+z^3+z^4-z^5+z^6-z^7 \\
&3-z-z^2-z^3+z^4+z^5+z^6+z^7 \\
&3+z+z^2-z^3+z^4-z^5-z^6+z^7 \\
&3-z+z^2+z^3+z^4+z^5-z^6-z^7,
\end{align*}
and the remaining polynomials for $k=1$ are yielded by the transformation
\begin{equation*}
B(z) \mapsto z^7\,B\Bigl(\frac{1}{z}\Bigr).
\end{equation*}
(We leave it to the reader to write down the corresponding Bell inequalities in 
traditional notation.)\hfill\ensuremath{\Diamond}\medskip

If $N$ is small, we can solve analogous problems also by brute force. 
But even for not too large $N$, this is hopeless since the number of Bell 
inequalities not only grows exponentially, but \emph{super}exponentially. 
With the tools provided in this paper, however, specific results can even be 
obtained for large $N$.

\section{Outlook}
So far nothing has been said about quantum violations, which can also be investigated in the 
presented context. Hereby, the maximal violation is always obtained for a 
generalized GHZ state \cite{ghsz,werner-wolf:01}. We also mentioned that by taking symmetry 
transformations into consideration, the set of ``essentially different'' Bell polynomials 
is reduced; so a closer look at the structure of this condensed set is interesting. 
And, of course, we can also study more general setups (and the corresponding quantum violations), 
where each measurement has more than two possible outcomes or each observer 
has the choice of more than two observables.

Apart from being of theoretical interest for the foundations of quantum mechanics, 
Bell inequalities are also of practical use in quantum cryptography (which is sometimes
more accurately termed quantum key distribution). A.\,K.~Ekert \cite{ekert:91} presented a variant of 
the method developed by C.\,H. Bennett \& G. Brassard \cite{benn:84}, according to which
an eavesdropper may be recognized by checking whether a certain Bell inequality holds. 
This approach was recently expanded by V.~Scarani \& N.~Gisin to quantum communication 
between $N$ partners \cite{scar:01}, where the security of this communication is again linked to 
violation of Bell inequalities.

\begin{acknowledgments}
Many thanks to Karl Svozil for his support and highly appreciated remarks,
and to \v{C}aslav Brukner and Marek \.{Z}ukowski for discussions related to this work. 
Special thanks go to Susanne Steinacher for linguistic support.
\end{acknowledgments}

\end{document}